**Title**

Fluorescence-enhanced photodynamic assessments for human breast cancer cell characterizations

Mahsa Ghezelbash[1]*, Batool Sajad[1]*, Shadi Hojatizadeh[2]

[1] Department of Atomic and Molecular Physics, Faculty of Physics, Alzahra University, Tehran, Iran

[2] Department of Biotechnology, Faculty of Biological Sciences, Alzahra University, Tehran, Iran

*Correspondence

Mahsa Ghezelbash, Department of Atomic and Molecular Physics, Faculty of Physics, Alzahra University, Tehran, Iran.
Email: m.ghezelbash@alzahra.ac.ir

Batool Sajad, Department of Atomic and Molecular Physics, Faculty of Physics, Alzahra University, Tehran, Iran.
Email: bsajad@alzahra.ac.ir

*((Mahsa Ghezelbash and Batool Sajad have contributed equally to this work.))*

**Abstract**

The primary and metastatic stages of BC consider different cancerous cell lines (MDAs), meaning the highest number of cells in this research (300,000) represents the metastatic stages of BC, and 50,000 represents the primary level of BC developments have been studied based on fluorescence-enhanced photodynamic characterizations. The ability to characterize the fluorescence caused by MDA with 50,000 cells compared to the dominant radiation of MDA with 300,000 cells is emphatic proof of the high potential of fluorescence technique in timely BC detections. 5-ALA induces an accumulation of protoporphyrin IX (PpIX) photosensitizer through a biosynthetic pathway, leading to red radiation of fluorescence measurements. The fluorescence signal reduction due to decreased cell viability and increased MDA's cellular death rate corresponds to the changes in lipid metabolism enzymes of MDAs cultured at different doses. Furthermore, statistical concerns have been studied using PCA multivariate component analysis to differentiate MDA cells administered by 5-ALA.

# 1    INTRODUCTION

Breast cancer, with an increased incidence of 12%, estimated new cases of around 2 million, and 20-70% positive margins, is the most frequently diagnosed malignant in women worldwide [1–3]. The treatments proposed are continually improving, reducing the mortality rate[3–5]. However, they are still causing many early and late side effects due to the weakness of diagnostic techniques, especially in the early stages, leading to the pre/post-cancer period as a chronic condition. Consequently, moving toward well-being leads to proposing fluorescence-based structures for breast cancer cellular metabolism characterizations to consider the efficiency of treatments[6–8].

Despite expensive and complicated VETs such as X-ray mammography [9,10], MRI [11–13], and PET/CT [14,15][14, 15], their inadequate accuracy in tumor residuals detection or considering cellular metabolism variations, especially in the early stages, are challenges facing this complication today [16,17]. VETs suffer from several constraints, such as large hardware, high operation costs, radio hazard risk, and disruption of the surgical workflow [18].

Various optical techniques, including optical coherence tomography (OCT) [19,20], Raman spectroscopy [21], and fluorescence lifetime imaging microscopy (FLIM) [22,23]. Hyperspectral imaging [24] have been investigated for their potential in breast cancer-characterizing cancerous breast cells, offering non-invasive approaches to obtain high-resolution images, analyze the molecular composition, study fluorescence decay rates, and detect spectral differences in tissue and cellular samples.

Fluorescence-based techniques by developing early detections could lead to a fully cured breast cancer disease because of their high sensitivity to alterations in cells' function, morphology, and micro-environment [25]. While a low resolution still hinders the application, this procedure exhibits an ideal probe targeting the breast tumor cells based on fluorescence high quantum yield. High sensitivity with reliable and fast measurements made that a qualified method, especially for middle/low-income communities.

This study hypothesizes that fluorescence-based approaches concerning photodynamic assessments hold significant potential for accurate discrimination and characterization of cancerous and noncancerous breast cells. By utilizing 5-aminolevulinic acid (5-ALA) as a probe, this approach can effectively characterize biomarkers metabolic changes and provide valuable insights into functional and morphological alterations associated with breast cancer progression. It has been proposed to aid the characterization of various malignancies by studying pathological changes [26]. Thus, fluorescence could yield information about breast cancer's physiological states from reliable key fluorophores (KF). Among those KFs, tryptophan, collagen, elastin, reduced nicotinamide adenine dinucleotide (NADH), and flavin adenine dinucleotide (FAD) are the primary native KFs relating to BC cellular metabolic and functional processes [26–28].

Fluorescence-enhanced photodynamic assessments aimed to demonstrate the technique's potential in distinguishing between cancerous and noncancerous cells based on spectral shifts, fluorescence intensity differences, and metabolic changes, which also investigated the impact of the 5-ALA prodrug on cancer cell characterization, photodynamic processes, viability, and morphology offer valuable information about the physiological states of breast cancer. This update aims to provide a comprehensive report on the efficacy of fluorescence-enhanced photodynamic examinations as a primer for oncology in studying breast cancer cells. With a high level of trustworthiness, the update delves into the conceptual and technical foundations of the fluorescence approach. It summarizes the practical and technical aspects of using the fluorescence-enhanced technique and the results obtained. Furthermore, it reviews the pathophysiological mechanisms involved and outlines the limitations to consider when applying this approach to breast cancer research.

The 5-ALA prodrug affects MDA viability, leading to a high rate of cellular death based on the applied doses of the prodrug due to its influence on intracellular transports [29]. The Reactive Oxygen Species (ROS) are produced in all cells because of normal metabolic processes, and only when their levels are significantly increased or decreased might disease be indicated. ROS could indeed be involved in cancer progression and can be used as a monitoring indicator that could cause genetic instability due to DNA damage or mutation load [30,31]. To detect even partial amounts of cultured MDA cells up to 50,000 indicates the high accuracy of the fluorescence-based approach, especially in early BC investigations. The apparent difference in the intensity of irradiated fluorescence for the highest amounts of breast cancer cell line (MDA) of 300,000 and other cell values spectral discrimination also proves the high sensitivity of the fluorescence technique in the study of BC.

The key fluorophore molecules (likely NADH and FAD) explore cellular metabolism changes [8,32] that arise from HFF transformations to BC cells (MDAs), causing discrimination between MDAs and HFFs. MDA's different number of cells (50,000-300,000) varies fluorescence emission according to their metastatic potential, showing four times stronger emission at around 500-510 nm than HFFs. The endogenous fluorophores or auto-fluorescent molecules in the cells or the surrounding environment were observed in all 5-ALA-induced MDA cells, such as Flavin, porphyrins, and lipofuscins [33,34].

Correspondingly, 5-ALA-administered MDAs significantly reduced cell viability and highly affected cell lines' morphology according to different cell numbers of 50,000-300,000 due to their different lipid contents [29].

## 2    MATERIAL AND METHODS

2.1  Cell lines

For fluorescence measurements, MDA-MB-231, human foreskin fibroblast cells (HFF-2) as "control" have been cultured with six different numbers of cells of 50,000, 75,000, 100,000, 150,000, 200,000, and 300,000. Human MDA-MB-231 breast cancer and HFF-2 cell lines were purchased from the Stem Cell Research Center, STRC (Tehran, Iran) [35]. All cell lines were cultured in Dulbecco's Modified Eagle Eagle's Medium (DMEM) supplemented with 10 percent fetal bovine serum (Gibco serum albumin) and 1 percent antibiotics seeded in two 6-well cell culture plates of DMEM and incubated for 24 h (%5 $CO_2$, 37°C, and 95% humidity). After incubation, the medium was removed and replaced with a fresh FBS.

At the second level of the experiments, 5-aminolevulinic acid hydrochloride (5-ALA) was dissolved in sterile distilled water.

5-ALA administrated MDAs with six different cell amounts were prepared at a concentration of 100 µM in DMEM medium with FBS for 24h at 37 °C in a standard 6-well cell plate was then removed from the wells by trypsin, dissolved at a 1.5 ml PBS.

Based on previous research and experimental considerations, the study set a 24-hour incubation time for 5-ALA with cells. Extending this time was challenging due to the large volume of the growth medium of about 3 ml and the difficulties in keeping the cells alive. After multiple tests, 24 hours was the optimal and longest feasible incubation period. Cell culturing, incubation, and fluorescence measurements also presented difficulties, making it more efficient to cultivate cells in a large volume. Hence, extending the incubation time beyond 24 hours was feasible.

2.2  Instrumentation

Two experiments were used to acquire the spectra from cultured MDAs and HFFs separately.

Once used, a commercial HORIBA Florolog fluorometer was a proof of concept to examine the optimal excitation wavelength and consider the efficiency of the fluorescence-based feasibility. The other was the main experimental setup, designed and calibrated using proper optical components to characterize cell lines.

Commercial HORIBA Florolog was an altered fluorescence spectrophotometer equipped with a xenon lamp, and the excitation was adjustable at adequate exposure times. A photomultiplier tube as a detection part allows the system to take accurate measurements.

Also, this system assembled an adjustable cuvette holder to hold the quartz cuvette. The delay between the lamp pulse and the detector is adaptable, enabling the collection of fluorescence or phosphorescence delays. The fluorometer system was also adjustable with different slit widths from 5 to 3000 µm. The lamp illuminates samples, excitable from 210 to 800 nm.

In the second (main) experiment, a sensitive homemade spectrometer can collect light in the wavelength range of around 210 to 880 nm with a spectral resolution of 0.7 nm and an integration time of 100 µs to 10s.

Its configuration is highly reliable and easy to use, especially discrimination of each sample based on spectral features.

The samples were exposed at the 405 nm continuous wavelength of the CNI laser diode, less than 0.5 mm square, offering almost a high efficiency and lifetime compared to lamp sources. The laser can create 1 to 200 mW output power to excite samples. Despite the outstanding quality of the laser beam and the high stability, a set of filters including two bandpass filters (390nm CWL, 25mm Dia, 40nm Bandwidth, OD 6, 86-348, Edmond Optics) tilted at 5-7 degree were use to narrow wide pass-band in the excitation leg. Then, using two UV-fused silica plano-convex lenses with 50 mm focal lenses shown as optics block (O), the paralyzed passing beam led to the center of the quartz cuvette embedded at the center of the holder, shown in the schematic figure of the schema at **Figure 1**.

UV fused silica plano-convex lenses were selected due to their unique features on fluorescence applications. Exposing them to the beam pass will not lead to excess fluorescent radiation from the lenses themselves and reduce the sensitivity of the measurements.

To capture the graphical morphology from a CMOS (E3ISPM, LaboQuip Digital Camera) camera coupled with an achromatic objective lens (VIS-20X-95-NA0, 300-800 nm) was used as shown in **Figure 1** schema.

In the end, fluorescence emission was collected and guided through the spectrometer's entrance using two 409 nm dichroic long-pass filters (Edmond Optics) tilted at 5 to 7 degrees, along with two 50 mm UVFused silica plano-convex lenses shown as collection optics block (CO) in **Figure 1**. The setup was also covered with a non-fluorescent box to prevent annoying environmental noise.

2.3 Statistical analysis

Statistical analyses and graphs were performed using MATLAB, Excel, OriginPro, and SPSS software. The software SPSS has been used to plot and interpret multiple comparisons. All experiments were prepared and analyzed numerous times, and the obtained data were presented as the mean ± SD.

# 3     RESULTS AND DISCUSSION

3.1 MDA pre-investigations

The birth of the research, mainly for timely BC detections, required collecting a detailed database using the studies carried out in this field until now.

It then conducted preliminary practical examinations using any equipment on the BC cellular changes and investigated using Florolog as a necessary part to evaluate fundamental ideas.

Considering the initial point of this research as the zero stage, this is regarded as a (-1) level, evaluating the initial feasibility of the investigation and the proof of concept.

MDAs prepared with six different numbers of cells of 50,000, 75,000, 100,000, 15,000, 200,000, and 300,000 have been carried out to qualitatively examine their collected fluorescence changes and the sensitivity of measurements.

As can be seen in **Figure 2**, MDAs and HFFs at six different numbers of cells of 50,000-300,000 were subjected to collecting fluorescence emission spectra (excitation/emissions at 405 nm/420-600 nm) using a standard spectrofluorometer (Horiba Scientific, Edison, NJ, USA) equipped with a quartz cuvette holder.

The fluorescence spectra were recorded from 1 ml of each cell line. The obtained data were corrected from the blank quartz cuvette and ambient. For 405 nm excitation, a predominant native fluorophore at around 420-480 nm and 480-530 nm regarded NADH and FAD, respectively, with an almost high quantum efficiency depending on the local environment and protein structures, shows the most significant spectral discrepancy, which could count as convenient to maximize cancer discrimination [8,28,32].

To highlight the spectral differences among MDAs with varying cell counts (from 50,000 to 300,000), we compared the obtained fluorescence excited at 405 nm at the prominent spectral characterized peaks of 420, 454, 510, and 556 nm (**Figure 3** and **Table 1**), highlighting the spectral differences among MDAs with different cells from 50,000 to 300,000.

The fluorescence-enhanced photodynamic assessments provide accurate discrimination between MDAs and HFFs using direct exposure to FS as a natural photosensitizer, which has been used at the forefront of cell metabolism investigations [36,37]. The HFF-2 cell line represents fibroblast cells with elongated and spindle-shaped morphology, which are the most abundant among connective tissue cells. The primary function of fibroblasts is the structural cohesion of connective tissue, which counts as a "control." Due to some features, such as being accessible to culturing, longer life, and the high structural similarity of these cells to the human breast cells, they were very efficient, especially in comparing the cellular viability or death of these cells with MDA cells applying 5-ALA prodrug. Likewise, 5-ALA-induced PpLX is a potent fluorescent agent [38,39] and a photosensitizer favored in malignant and premalignant cells, making it an excellently targeted probe in photodynamic considerations [40–42].

Reactive Oxygen Species (ROS) are by-products of aerobic metabolism and can act as signaling molecules to participate in multiple regulation of biological and physiological processes [43]. They Hydrogen Peroxide ($H_2O_2$), superoxide anion ($O_2^-$), hypochlorous acid (HOCl), singlet oxygen ($^1O_2$), and hydroxyl radical ($\cdot OH$), and act as second messengers in cell signaling, and are essential for various biological processes in normal and cancer cells [43]. The occurrence, growth, metastasis of tumors, and even the apoptosis, necrosis, and autophagy of tumor cells are all closely related to ROS [44].

5-aminolevulinic acid (5-ALA) is a non-fluorescent prodrug that can be metabolized into fluorescent protoporphyrin IX (PpIX) in cells often used in photodynamic therapy (PDT) for cancer treatment. In this process, 5-ALA is administered to the patient, accumulating in the cancer cells. When exposed to a specific wavelength of light, the PpIX is activated and produces ROS, which can cause cell death.

MDAs show an increased NADH fluorophore significantly at around 454 nm and FAD at 510 nm emissions compared to each other could be related to the different other fluorophores most likely [8,45].

The minimum number of cells is 50,000 (early BC), which is targeted to evaluate the detection ability of the fluorescence platform preclinical assays. It shows the comparison of the lowest fluorescence intensity compared to 300,000 (advanced BC) on different wavelengths of 420, 545, 510, and 556 nm, admitting that any MDA cellular changes (metabolism) [37,46] could happen fluorescence variations, clarify fluorescence's ability to detect, especially in early BC.

3.2 MDAs and HFFs experimentation

It could also be considered a comparison of the fluorescence variation profiles regarding MDAs and HFFs through increasing the numbers of cells in the emissions areas of 475-545 nm under the 405 nm excitation wavelength given (**Figure 4**). Exposing samples at 405 nm excitation using the designed fluorescence platform (**Figure 1**) ignited Fluorescence-enhanced photodynamic assessments.

It can be seen that MDAs exhibit fluorescence emission almost four times more substantial than that of HFFs at the same wavelength region belonging to 300,000 cells, as **Figure 4** (a) and (b). In contrast, HFFs exhibit equal or minor intrinsic fluorescence signal variation rates (c and d). Note that rapid measurements were recommended in this research since there was a higher possibility for cell destruction in case of extended measures due to being cells longer out of standard storage conditions.

The time to measure a complete set of prepared cells (6 different samples each day) needed at least 20 minutes.

Therefore, the obtained sets were averaged accordingly and needed at least three measurements repeated on different working days, which helped us compare the data each day with a minimum error.

However, there is a noticeable difference in intensity between the fluorescence data of MDAs and HFFs, especially around the 500 to 530 nm emission area with four distinct regions at around 480, 500, 510, and 530 nm that could be arising from essential coenzymes [7,8]. Based on existing literature, NADH and FAD could be the most blocking coenzymes emitting around 480 to 540 nm and highly depend on their cells' lipid metabolism enzyme changes, corresponding to fluorescence variation [6,47–49].

3.3 The 5-ALA authority in the MDA's cellular viability and morphology

The potential death mechanism induced by 5-ALA prodrug and its treatment effects has been investigated. 5-ALA prodrug has an essential impact on the BC cell morphology that has been evaluated (**Figure 5** and **6**). The phase-contrast images were captured using a camera setup consisting of a microscope (XDS-1B inverted biological lab microscope) attached with a CMOS camera (Toupcam-U3CMOS-10000KPA, 10 MPix) and a proper camera lens in which the cultured cells before/after incubation adding ALA directly from a 6-well cell plate was illuminated perpendicularly using a LED white light source. The attached microscope then collects the reflected light with a standard objective lens with a focal distance of f = 200mm. Captures of a CMOS camera monitored the cellular morphology of MDAs before and after 5-ALA administrations for about 24 hours. As shown in Figs. 5(a), MDAs Inherently display polyhedral morphology, causing them with low metastatic potential [50,51]. While cell treatment 5-ALA-administered MDAs accumulates, cells appear more rounded than untreated MDAs **Figure 5** (b).

The ability to reveal the decreasing fluorescence signal caused by the 5-ALA-affected MDA cell lines is from the other side stepping forward through photodynamic considerations of BC characterizations.

That is an efficient approach to characterizing BC, showing a clear difference between 5-ALA-assisted and not-assisted MDAs in spectral areas of around 480-540 nm, as shown in **Figure 6** According to the cellular shape and fluorescence intensity of spectra, 5-ALA-assisted MDAs (a-f) show lower fluorescence intensities than untouched 5-ALAs.Our results complement that 5-ALA-assisted MDAs lead to a higher rate of cellular death and deformation rate, which decreases the fluorescence obtained[52,53]. This effect is because absorbing 5-ALA by BC cells and exciting at 405 nm, making 5-ALA produce oxygen radicals, leading to MDAs cellular death [7]. The fluorescence spectra of the 480-540 nm spectral areas differ slightly from the ALA-assisted MDAs, especially at the lowest number of cells from 50,000(a) than the maximum number of cells (f).

Enhancing a higher visualizing of the spectra variability, about 0.01 ml sodium fluorescein (SF) as a fluorescent dye [37] with a molecular weight of 376 g/mol has accumulated on each MDA with a different cell number of 50,000-300,000 through 6 different standards 1 ml micro-tube. Then, each sample was vortexed for 30 seconds, which prolonged exposure to 1ms of the activating 405 nm excitation marked obtained fluorescence assays with further implications potential.

The results demonstrate the potential of a fluorescence-based technique in distinguishing between cancerous and noncancerous breast cells based on spectral shifts, fluorescence intensity differences, and metabolic changes associated with tumor initiation and progression. Our objective was to assess the overall diagnostic capability of the fluorescence-based technique in characterizing breast cancer cells. The varying numbers of cells were utilized to investigate the impact of cell concentrations and characterization potential of 5-aminolevulinic acid (5-ALA), a prodrug used in fluorescence-enhanced photodynamic assessments By culturing breast cancer cells (MDAs) and normal human cells (HFFs) with different numbers of cells (ranging from 50,000 to 300,000), we aimed to evaluate how cell concentrations influence the uptake and subsequent fluorescence characteristics of the 5-ALA.

While higher cell numbers are expected to decrease the uptake per individual cell, studying a range of cell concentrations allows us to assess the overall fluorescence signal and its detection implications. The objective was to understand the relationship between cell concentrations, 5-ALA, and the resulting fluorescence emissions. This information is crucial for determining the optimal conditions and cell concentrations for a reliable and effective fluorescence-based solution. Administering 5-ALA induces PpLX directly into cultured cells and, after incubation, accumulates PpIX photosensitizer through a biosynthetic pathway. Based on the studies, it is expected that applying 5-aminolevulinic acid (5-ALA) causes the production of another biomarker named protoporphyrin IX (PpIX) as a part of fluorescence emissions, measurements sensitivity to cell detections and characterizations. Accordingly, it is supposed to appear at a peak around 620 and 634 nm (a reddish fluorescence), applying 405 nm light.

In comparison, the main emission interval (in this work) was around 420 to 600 nm (**Figure 2**) and 470-545 nm (**Figure 6**); hence, the fluorescence emission of PpIX is removed. It could be known to be pH-related [54,55] or link pathological status, and the gene expressions in vivo rely on a different aggregate of PpIX [56–58]. Also, other influential factors, such as temperature, incubation time, and added glucose of the culturing medium, influence the relative porphyrin concentrations, causing a faster photo-bleaching rate of the Pp spectrum [54,59,60].

3.4 Multivariate cell lines analysis

Principal Component Analysis (PCA) was used to objectively differentiate between the MDAs, HFFs, and the 5-ALA-administered MDAs.

**Figure 7** shows the PCA of 10 different measurements discontinuously for the principal components 1 and 2 in a plane. The scattering of the data points, which are a hollow blue triangle for HFFs, open black circles for MDA-(5-ALA), and solid red circles for MDA+ (5-ALa), visualizes the possibility of differentiating between them and the reproducibility from one measurement to another.

PCA was done on ten sets of cultured MDAs with a different number of cells of 50,000, 75,000, 100,000, 150,000, 200,000, and 300,000 during a similar cultural process separately. The fluorescence spectra of the MDAs and HFFs show intensity peaks at around 480 to 540 nm. Principle components one (PC1) and two (PC2) represent the data set from the fluorescence spectra excited at 405 nm. All experiments were repeated at least three times (n=3). The obtained data was presented as the mean ± standard deviation.

The higher concentration of hollow blue triangle for HFFs rather than MDA-(5-ALA) indicates that the obtained fluorescence spectra of HFF are reproducible, irrespective of measurements. The clear separation of the HFF data (triangles) and from the MDAs (circles) indicates that the obtained fluorescence spectra of HFF can be separated from MDAs. It is also considered that the dispersion of the data in the case of MDA+(5-AlA) has more similarity to the HFFs (with the first principal component 84.99% of the total variance in the fluorescence spectra of **Figure 7** (b).

While the PCA graph displays variability, which is anticipated due to the inherent biological differences in cell lines, even under similar culture conditions, this variability stems from factors such as genetic drift [61], variation in cell culturing conditions [62], genomics variations or epigenetic modifications [63,64], PCA refactors [65] and cell cycle stage disparities at the time of analysis. The number of cells can influence the intensity of the PCA graph, but it can also introduce variation if the cells are at different growth or confluence stages.

In essence, the PCA graph's variation results from the multidimensional nature of the data, which may not be fully represented in other graphs. The PCA graph offers a multidimensional data perspective and is not designed to differentiate between individual groups. Instead, it provides a comprehensive view of the overall data structure and relationships. The overlapping groups in the PCA graph signify the biological similarities among the cell lines and the subtle effects of the treatment.

## 4 CONCLUSIONS

This research highlights the high potential of fluorescence-based approaches for accurate discrimination and characterizations between cancerous and noncancerous cells based on biological changes associated with tumor initiation and progression as the main idea of Fluorescence-enhanced photodynamic assessments.

Noncancerous dermal fibroblast cells (HFFs) and cancerous cells (MDAs) with primary (50,000) up to advanced (300,000) metastatic stages of BC were growing under standard laboratory situations. In summary, using different cell numbers in the study may serve to investigate the relationship between cell concentration and fluorescence emission and provide insights into the behavior of cancerous cells at various stages of breast cancer. However, it is essential to consider the potential impact of cell concentration on the uptake of ALA and interpret the results accordingly for future work.

When 5-ALA is administered, it accumulates in cancer cells and is metabolized to protoporphyrin IX (PpIX). When PpIX leads to cell death. However, the specific dose of ROS that can induce cell death can vary depending on the type of cell, the cellular environment, and other factors. It is also important to note that ROS can cause cell death but plays a vital role in normal cellular functions. The metabolic changes of the cells under the influence of the induced (bio) chemical factor, such as 5-ALA, validate its biological activity based on spectral shifts, and fluorescence intensity differences are keys to characterizing BC cells for pre/post-clinical applications.

Photo-dynamical reactions of the fluorophore probes after 5-ALA registering and fluorescein sodium (FS) administration have also been studied, principally considering the cellular viability and visualization booster. Thus, after 24 h of 5-ALA stimulation, the MDA cells manifested significant cellular viability based on their morphology changes. The efficacy of 5-ALA prodrug on the higher number of MDA cells (200,000) in treating advanced breast cancer has already been revised in the literature.

It was considered that after 5-ALA administration to MDAs, the cells are rapidly metabolized, owing to changes in the activity of enzymes and highly affecting their BC cell viability, decreasing fluorescence emission.

The results also show a particular aspect of 5-ALA prodrug on MDAs, suppressing BC cells spreading to the surrounding areas, especially the axillary lymph nodes, preventing their progress, which can significantly affect cancer treatment. Comparing the microscopic images of MDA cells assisted with and without 5-ALA prodrug on the morphology of the cells could claim this approach as photodynamic procedures showing up as the other sides of the obtained results.

**ACKNOWLEDGMENTS**

This work was funded by the Center of International Science and Technology Cooperation (CISTC) and was supported by Alzahra University in Tehran, Iran.
The authors thank Dr. Reihaneh Ramezani, AP of the Nanobiotechnology research group from the Women Research Center at Alzahra University, for her scientific and technical assistance with MDA‐MB‐231 and HFF cell line preparations.

**AUTHOR CONTRIBUTIONS**

M. Gh. and B. S. contributed equally to this paper. Sh. H. from the biological group collected and prepared the total MDA and HFF cell lines. Sh. H. also participated in the interpretation of results and data finalization.

**CONFLICT OF INTEREST**

The authors declare that they have no conflict of interest.

**DATA AVAILABILITY STATEMENT**

The corresponding authors' data supporting this study's findings are available upon reasonable request.

**FIGURES AND TABLES**

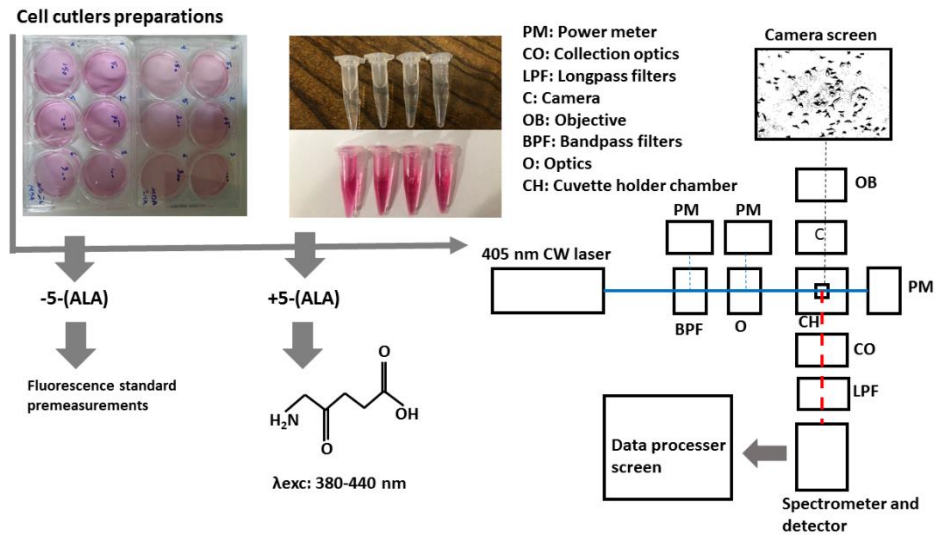

**FIGURE 1.** [Schematic setup for cell lines fluorescence characterizations.]

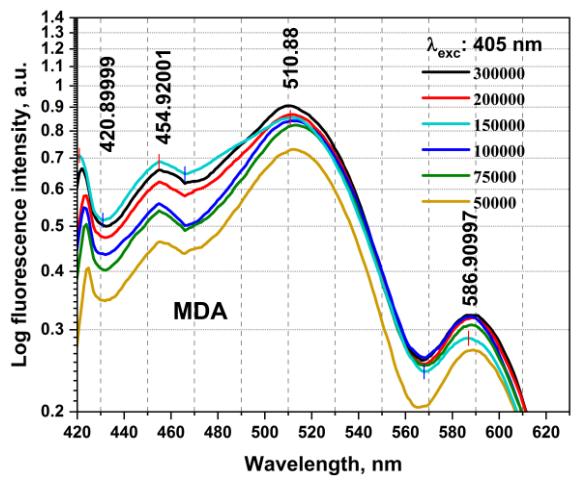

**FIGURE 2.** [The fluorescence spectra registered from the excised MDA specimen were obtained from different cellular numbers of 50,000, 75,000, 100,000, 150,000, 200,000, and 300,000. Each measurement was tested in three independent experiments (n=3)]

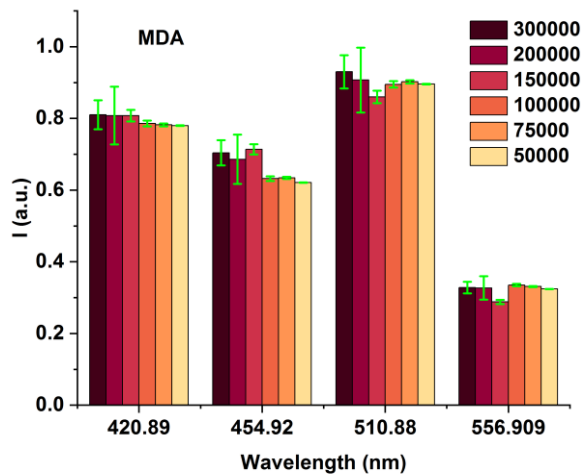

**FIGURE 3.** [Bar graphs of different MDA cell numbers of 300,000, 200,000, 150,000, 100,000, 75,000, and 50,000 fluorescence intensity correspond to 420, 454, 510, and 554 nm emission wavelengths under 405 nm excitation.]

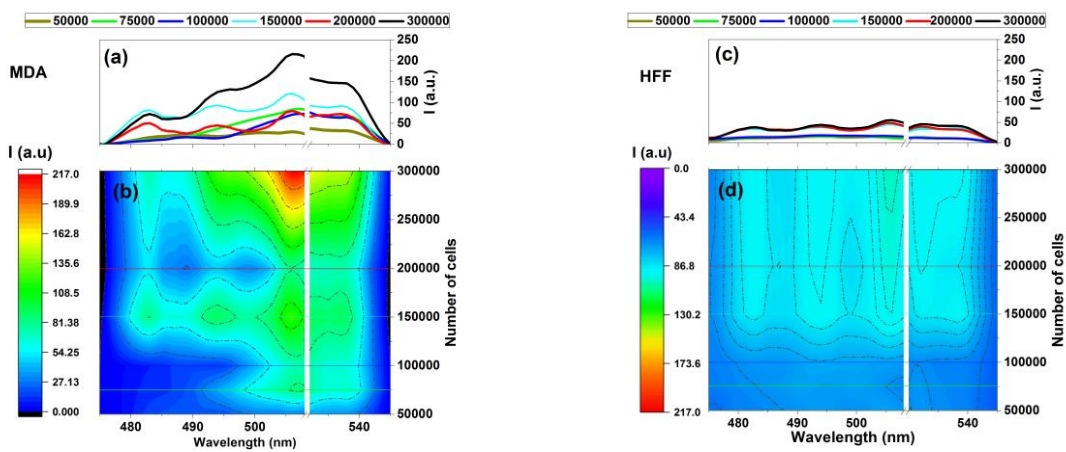

**FIGURE 4.** [Fluorescence intensity vs. the number of cells for (a and b) MDAs and (c and d) HFFs at an excitation/emission of 405nm/ 454-545 nm correspond to the fluorescence of NADH at different numbers of cells of 50,000, 75,000, 100,000, 150,000, and 200,000. Each measurement was tested via three independent experiments (n=3)]

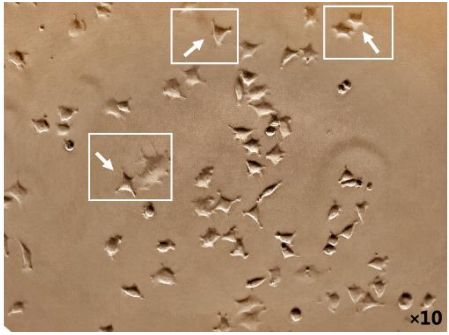
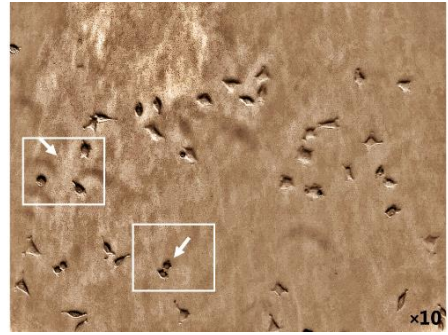

**(a)** **(b)**

**FIGURE 5.** [Cellular morphology of control MDA human breast cancer cells monitoring at 50,000 number of the cell (a) before and (b) after 1mM of 5-ALA treatment for 48 h by phase-contrast photos.]

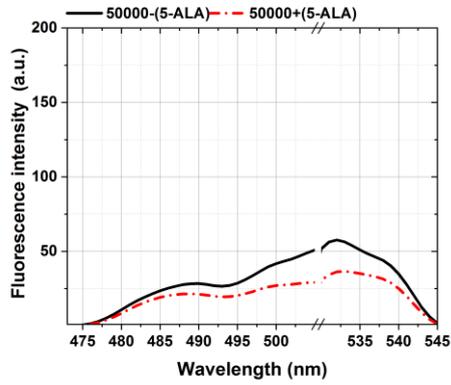
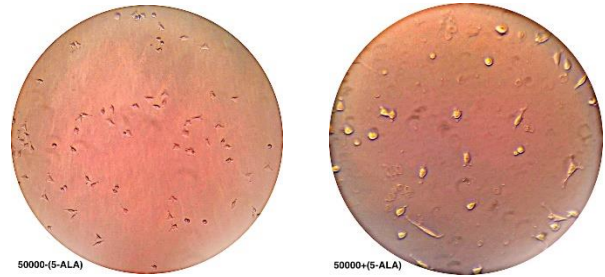

**(a)**

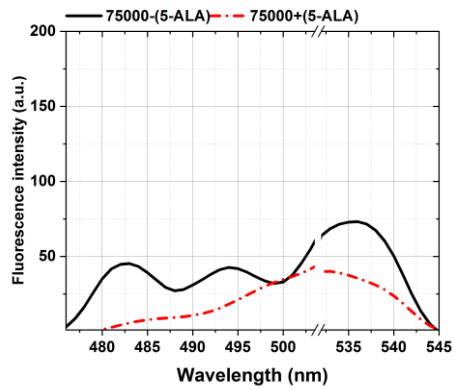
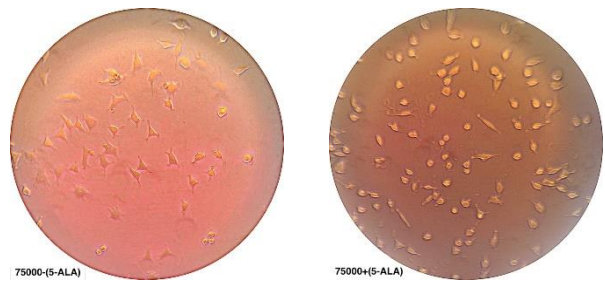

**(b)**

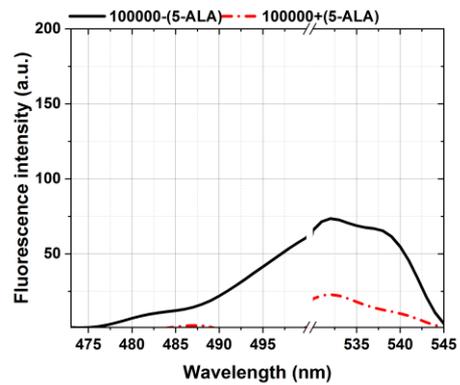
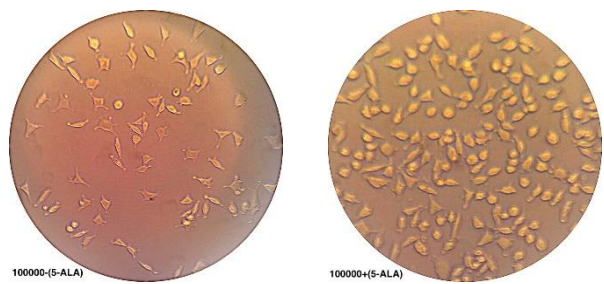

**(c)**

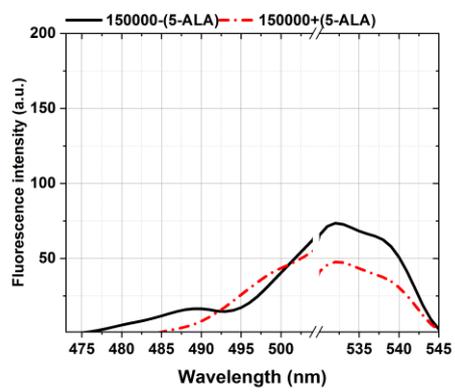
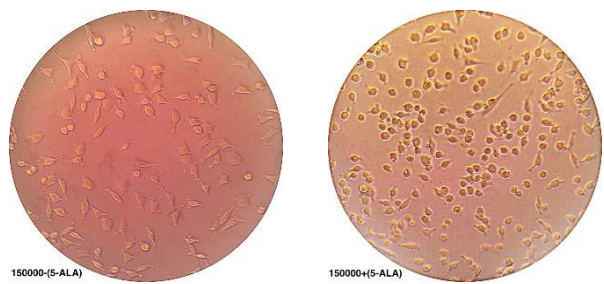

**(d)**

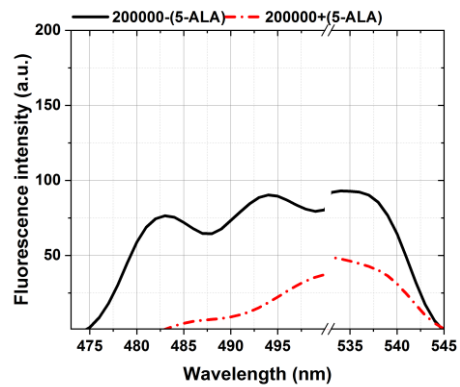
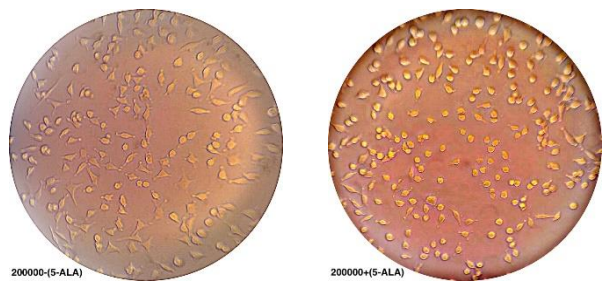

**(e)**

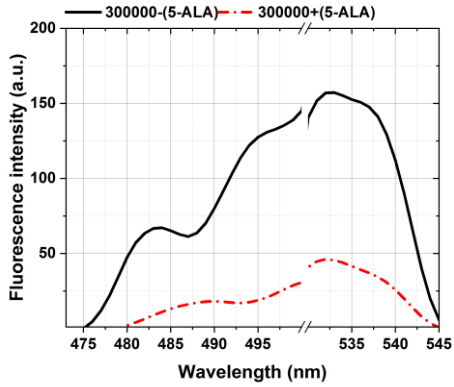
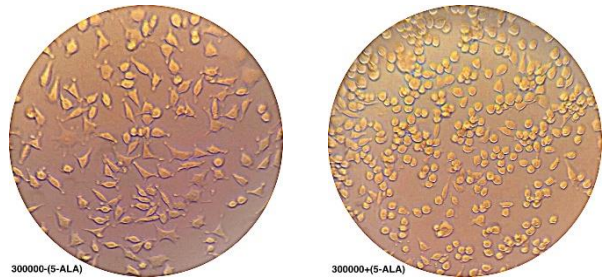

**(f)**

**FIGURE 6.** [Comparison of the fluorescence spectra (left plots) and different MDAs cellular shapes (right photos) from different MDAs number of cells (50,000-300,000), assisted with 5-ALA and without excited at 405 nm excitation. Scale bars, 100 mm. Each measurement was tested in three independent experiments (n=3).]

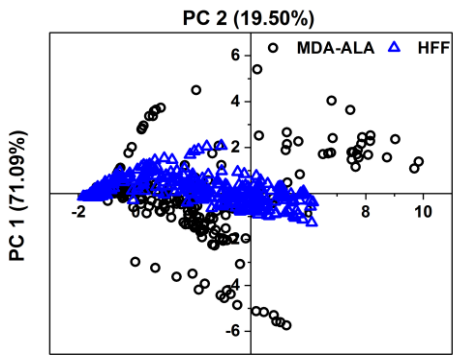
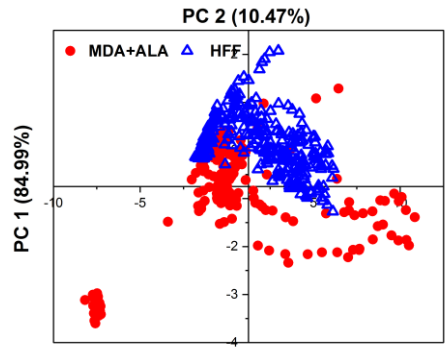

**(a)** **(b)**

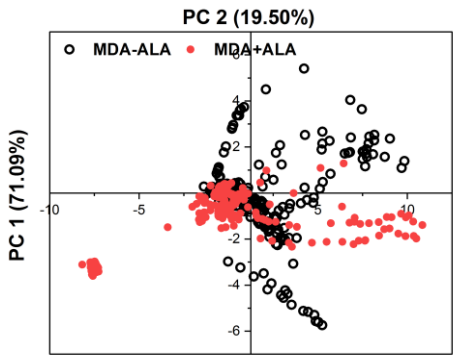

**FIGURE 7.** [Comparison of the PCA results between (a) MDA (without ALA) and HFF, (b) MDA (with ALA) and HFFs, and (c) MDAs assisted with and without ALA at 405 nm excitation. Data was repeated over ten times in the calculation.]

**(c)**

TABLE 1. [Wavelength specifications.]

| Index    | Beginning x [nm] | Ending x [nm] | FWHM [nm] | Central wavelength [nm] | Height |
|----------|------------------|---------------|-----------|-------------------------|--------|
| 420.9 nm | 409.8            | 30.9          | 15.4      | 420.9                   | 0.7    |
| 454.9 nm | 430.9            | 465.8         | 32.9      | 454.9                   | 0.6    |
| 510.8 nm | 465.9            | 567.9         | 80.6      | 510.9                   | 0.8    |
| 586.9 nm | 567.9            | 643.9         | 48.4      | 556.9                   | 0.2    |